# STRUCTURAL PECULIARITIES, MINERAL INCLUSIONS AND POINT DEFECTS IN YAKUTITES – A VARIETY OF IMPACT-RELATED DIAMOND

Andrei A. SHIRYAEV[1,2], Anton D. PAVLUSHIN[3], Alexei V. PAKHNEVICH[4,5], Ekaterina S. KOVALENKO[6], Alexei A. AVERIN[1], Anna G. IVANOVA[7]

[1] A. N. Frumkin Institute of Physical Chemistry and Electrochemistry RAS, Leninsky pr. 31 korp. 4, Moscow 119071, Russia.

[2] Institute of Geology of Ore Deposits, Petrography, Mineralogy, and Geochemistry, Russian Academy of Sciences, Moscow, 119017 Russia.

[3] Diamond and Precious Metal Geology Institute, Siberian Branch of RAS, Lenin pr. 39, 677000, Yakutsk, Russia.

[4] Paleontological Institute RAS, Profsoyuznaya str. 123, Moscow 117997, Russia

[5] The Frank Laboratory of Neutron Physics, JINR, Dubna, Russia, 141980

[6] NRC ''Kurchatov Institute'', Kurchatov square 1, Moscow 123182, Russia.

[7] Shubnikov Institute of Crystallography FSRC "Crystallography and Photonics" RAS, Leninsky pr. 53, 119333, Moscow, Russia

**Abstract.** An unusual variety of impact-related diamond from the Popigai impact structure - yakutites - is characterized by complementary methods including optical microscopy, X-ray diffraction, radiography and tomography, infra-red, Raman and luminescence spectroscopy providing structural information at widely different scales. It is shown that relatively large graphite aggregates may be transformed to diamond with preservation of many morphological features. Spectroscopic and X-ray diffraction data indicate that the yakutite matrix represents bulk nanocrystalline diamond. For the first time, features of two-phonon infra-red absorption spectra of bulk nanocrystalline diamond are interpreted in the framework of phonon dispersion curves. Luminescence spectra of yakutite are dominated by dislocation-related defects. Optical microscopy supported by X-ray diffraction reveals the presence of single crystal diamonds with sizes of up to several tens of microns embedded into nanodiamond matrix. The presence of single crystal grains in impact diamond may be explained by CVD-like growth in a transient cavity and/or a seconds-long compression stage of the impact process due to slow pressure release in a volatile-rich target. For the first time, protogenetic mineral inclusions in yakutites represented by mixed monoclinic and tetragonal $ZrO_2$ are observed. This implies the presence of baddeleyite in target rocks responsible for yakutite formation.

## INTRODUCTION

Several types of terrestrial impact-related natural diamonds are known. The most abundant type is represented by small, usually less than 1 mm in size, platy grains found in many large impact craters in the world likely formed during direct graphite-diamond transition in a shock wave, thus giving a name "apographitic" diamonds. The largest deposit is at the Popigai impact crater (see Popigai impact structure… 2019), but similar diamonds, although in smaller amounts, were found in many craters worldwide (Val'ter et al., 1992; Vishnevsky et al., 1997), e.g., in suevites from the Ries (Rost et al., 1978), in drill cores of the Belilovka (Tsymbal et al. 1999) and Puchezh-Katunki (Raikhlin and Shafranovsky 1999; The Puchezh-Katunki impact structure, 2020) impact structures to name a few; in some cases erosion leads to the formation of placers with impact diamonds. An interesting variety of impact-related diamond-containing material, possibly formed by impact alteration of coal or similar carbonaceous matter, was described in the Kara astrobleme (Northern Russia) and termed togorites (Yezersky 1986) or "apocoal" diamonds (Shumilova et al., 2018).

Yakutites, for historical reasons sometimes called "carbonado with lonsdaleite" or diamond variety XI (Orlov and Kaminskiy, 1981), represent an abundant, but perhaps outside of the ex-USSR less known variety of impact-related diamonds, first found in placers of North Eastern Siberia in the 1960s. They are large (up to ~1 cm), usually black or brownish polycrystalline diamonds, generally of irregular shape. They are found in placers of rivers draining the Popigai structure both within its boundaries (Popigai impact structure… 2019) and in an extensive region to the north-east. The most distant finds are as much as 500-600 km from the impact crater (Grakhanov 2005). Yakutites are often ascribed to eroded fine-grained polymict breccia (coptoclastites) or distant impact ejecta. Results of spectroscopic and structural studies of yakutites are summarized in (Galimov et al., 1980; Kaminsky et al., 1985; Bokii et al. 1986; Kharlashina and Naletov 1991); more recent advances were described in (Titkov et al. 2004; Ugap'eva et al., 2010; Yelisseyev et al., 2016a, 2018; Ohfuji et al. 2017). Note that not all studies fully concur with the impact origin of yakutites (e.g., Titkov et al. 2004 and refs. therein), but the majority of such works reject an interpretation of the Popigai structure as an impact-related one.

Despite a rather significant number of contributions, many issues related to yakutites remain debatable. In particular, it is not yet clear which factors lead to the formation of markedly different types of impact diamonds in the Popigai structure (small tabular diamonds found *in situ* in shocked rocks and yakutites) and whether yakutites are genetically linked to Popigai or to another, yet unknown impact structure. In this work, we describe results of investigation of several yakutite samples using high resolution optical microscopy, photoluminescence and infra-red spectroscopy, absorption and phase contrast X-ray tomography and radiography. We also provide information on primary inclusions in yakutites.

## SAMPLES AND METHODS

A collection of more than 200 yakutite samples recovered from alluvial sediments of the Anabar river diamondiferous region was studied. On average they are 4-5 mm in the largest dimension with masses between 0.08-1.12 g. The surface is strongly corroded; secondary minerals form patchy films and fill some of cracks and/or surface depressions. The coloration is uneven; the corroded domains are colored by iron hydroxides and vary from brown to dark or yellowish brown or black. They can be marginally translucent. Color of fresh fractures varies from dark grey to black.

Four samples were selected as representative of different morphological types and studied in detail in this work. A general view of the samples is shown in Fig. 1. Surface morphology as well as some other mineralogical features of these samples were discussed in (Ugap'eva et al., 2010). In the current work morphological peculiarities of the samples were studied using 3D optical microscopy employing a Keyence VHX-1000 device at magnifications up to 5000×.

Raman and photoluminescence spectra of the samples were recorded at room temperature using several excitation wavelength – 405, 488 and 633 nm (only Raman for the later) using inVia Reflex (Renishaw) and LabRam (Jobin Yvon) spectrometers. Raman spectra were recorded using a 1200 lines/mm grating; for photoluminescence (PL) a 600 lines/mm grating was used. In both cases, Peltier-cooled CCD detectors were used. Spectra were acquired using 50× and 100× objectives with a NA of 0.5 and 0.8, respectively. Laser spot size was between 1-5 μm depending on the laser, objective and eventual defocusing. Accuracy and resolution of lines in the Raman spectra are ±1 cm$^{-1}$ and ±2 cm$^{-1}$, respectively, and, for photoluminescence spectra are ±0.1 nm and ±0.3 nm. Laser power on the sample did not exceed 0.2 mW as measured using a LaserCheck device (Coherent); repeated control of spectra consistency was performed. Wavenumber calibration was performed using a Si reference sample and a high-quality CVD diamond. Infra-red spectra in reflectance geometry were collected using an AutoImage microscope attached to a SpectrumOne FTIR spectrometer (Perkin Elmer), with apertures between 50 and 100 μm. Infra-red spectra were recorded in transmission geometry in IR-transparent spots present in some samples. No special sample preparation except surface cleaning was applied. The thickness of the sampled region is between 1-2 mm. For detailed investigation of the $CO_2$ bands nitrogen-purged iN10 (Nicolet) IR microscope was used; the aperture size was between 50 and 150 μm. Spectral resolution was 2 cm$^{-1}$.

Initial assessment of the internal imperfections was performed using the mapping mode of an XGT-7200V X-ray fluorescence microscope (Horiba) employing a Rh anode operated at 50 keV and 1 mA. This device allows not only the acquisition of X-ray fluorescence spectra, but also provides radiographic images recorded using a scintillation screen underneath the sample. Importantly, only elements heavier than Na can be detected.

Subsequently, details of the internal structure of the samples was studied using X-ray tomography. Absorption contrast tomography was performed using a laboratory-based SkyScan 1172 setup operating at 100 μA - 49 kV, CCD image pixel size – 7.16 µm, angular step – 0.7°, rotation – 180°. Slice reconstruction was performed using Nrecon 1.6.4.1 software. White beam both with and without Al filters was used; spatial resolution was 10 µm. To reveal weakly absorbing details, in-line phase-contrast synchrotron tomography at the MEDIANA beamline of Kurchatov synchrotron source (Podurets et al., 2012) was involved. Synchrotron radiation (SR) from a bending magnet was monochromatised using a Si (220) crystal; the energy was 20 keV. The source-sample distance is 15 m; detector - sample distance - 32 cm (2 cm for absorption contrast); vertical source size is ~150 μm. The sample was rotated about the vertical axis with 0.05-0.1° steps. Projections of the sample were recorded by a position-sensitive detector with a GdOS(Tb) scintillation screen and a CCD 4008×2670 array. The beam size on the sample was $3 \times 35$ mm$^2$ and the spatial resolution was 28 µm. The exposure time of one image was 3.1-12 s. For initial processing ImageJ software (Schneider et al., 2012) was used; tomographic reconstruction was performed with the Octopus Imaging package (Dierick et al., 2004).

X-ray diffraction was performed at Xcalibur EOS Agilent Technologies (Oxford Diffraction) and at XtaLAB Synergy-DW (Rigaku) single crystal diffractometers equipped with CCD detectors. Monochromatic Mo Kα radiation was used, the diameter of the collimators was 0.2 and 0.5 mm. The X-ray beam was pointed to an apex of the specimen, which was rotated around vertical axis with a 0.1-0.5° step.

Two samples were laser-cut using tomographic and radiographic images for aiming and gently polished to expose mineral inclusions. During polishing utmost care was applied to avoid excess heating and was stopped as soon as the inclusions were reached. Quality of the polish of the yakutite matrix itself is highly variable due to the above limitations and due to absence of "soft" polishing directions.

## RESULTS

*Morphology of the studied samples*

Yakutites generally inherit the morphology of precursor graphite aggregates. Although the graphite to diamond transition induces deformation and subsequent oxidation at high temperatures may distort initial shape of graphite, in case of impact diamonds these factors are usually manifested as tiny surface features only (e.g., Kvasnitsya and Wirth, 2013 and refs. therein).

According to the classification of the external morphology proposed in (Ugap'eva et al., 2010) the studied yakutite samples belong to three main types (see Fig. 1).

The sample 401 belongs to a massive morphological type. One side is highly reflective and looks "molten". Other sides can be described as epigenetic uneven fractures implying that the specimen is a part of a larger piece. For natural graphite similar rounded shapes and botryoidal aggregates are known and are characterized by concentric zoning or fibrous internal structure.

The sample 406 preserves tabular morphology of graphite precursor, which was composed from parallel stepped flakes. The specimen also shows epigenetic chipping. Pinacoidal morphology (Fig. 1c in Ugap'eva et al., 2010) is typical for impact diamonds from Popigai (Tsymbal et al., 1999).

The sample 407 is the first example of a rare morphological variety of yakutite (Fig. 1e in Ugap'eva et al., 2010) and was formed from columnar, partly fibrous graphite. The precursor graphite was plastically highly deformed, possibly prior to the impact event.

The sample 408 was formed by transformation of several tabular graphite crystals. On one side three principal individuals are distinguished; another side is epigenetically chipped. Optical microscopy reveals fine peculiarities of surface relief of the graphite precursor with sizes from 5-10 up to 50 μm (Fig. 2) and morphological details of graphite flakes. On fractured faces steps are observed. Features of plastic deformation, cracks along the (0001) plane of perfect cleavage and trigonal hatching on (0001) common for natural graphite are resolved. All studied samples show matt relief due to numerous submicron depressions, probably resulting from epigenetic dissolution (Fig. 2e,f).

*Single crystalline diamond grains*

One of the most remarkable result of the current investigation of yakutite samples using conventional and 3D optical microscopy is the observation of large (up to 20 μm) diamond grains, which closely resemble single crystals; even (quasi)octahedral morphology is sometimes apparent (Fig. 2a-d). Image filtering using the "relief" function of the Keyence software was applied to enhance edges (high frequency signal) and other similar imperfections (fig. 2b). Large continuous domains are observed on the filtered images, confirming that these grains are indeed micron-sized single crystals. To provide solid proof of single crystal character of the observed grains, X-ray diffraction patterns were acquired using both "white" and monochromatic radiation. As expected, the patterns of yakutite samples show strong Debye rings arising from textured material and are similar to those reported in many earlier works (e.g., Galimov et al., 1980, Kaminsky et al, 1985, Yelisseev et al. 2018). However, a careful search revealed tiny, but distinct spots superposed on the textured Debye rings (Fig. 3). Occurrence of the spots is a solid proof of presence of diamond single crystals with sizes exceeding few microns embedded into nanocrystalline yakutite matrix. Occurrence of such grains is discussed in some detail in the Discussion section.

*Infra-red spectroscopy*

In some samples, in particular, in specimen 401, it was possible to find regions sufficiently transparent in IR spectral range to record useful spectra. Typical IR spectra of yakutite are shown in Fig. 4. As expected, the two-phonon region is dominated by broadened and distorted absorption bands of the diamond lattice, see below. The one-phonon region lacks features typical for nitrogen-related defects and the most prominent feature is a "triangular" band described earlier for impact diamonds (Klyuev et al. 1978, Yelisseyev et al., 2016a,b) and synthetic bulk nanocrystalline diamonds (Shiryaev et al., 2006). The main features of this band are at ~1030, 1225 (main peak) and 1326-1329 cm$^{-1}$. This band is tentatively assigned to defects related to stacking faults, perhaps, those similar to lonsdaleite. Such as assignment is supported by the observation of this band in diamond single crystals, heavily deformed at static high pressure – high temperature conditions (Shiryaev et al., 2007). Whereas solid proof is yet absent, a feature at 1326-1329 cm$^{-1}$ might correspond to a Raman frequency of heavily distorted and/or nanocrystalline diamond. In macroscopic diamond the Raman frequency is 1332 cm$^{-1}$, but in nanodiamonds and at least in some impact-related samples it is down-shifted. The one-phonon absorption at the Raman frequency is forbidden in ideal diamond lattice, but any defect lowering local symmetry may lead to the appearance of a corresponding IR peak (Birman, 1974). Some other weak bands are also present and most likely they are related to secondary surface-bound mineral phases and are not analyzed in detail. The three phonon region is influenced by light scattering making analysis ambiguous, but it is likely that the spectral envelope differs from that in ideal diamond. Besides the contamination–related C-H vibrations at ~2900 cm$^{-1}$, no features due to lattice-bound hydrogen (e.g., the 3107 cm$^{-1}$ peak) are observed.

The IR two-phonon absorption in yakutite differs from reference single crystal diamond in some detail, namely, the narrow bands at 1970 and at 2160 cm$^{-1}$ are smeared and their relative intensities are small. Whereas these changes might seem to be an artefact due to light scattering and other effects influencing background, examination of spectra of yakutite and other nanocrystalline diamonds shown in other papers (Shiryaev et al. 2006; Yelisseyev et al. 2016a) indicates that the effect is real. Many different phonons contribute to the two-phonon IR absorption. The principal "missing" features are assigned to two-phonon bands $\Sigma_1 A + \Sigma_3 A$ and $\Sigma_2 O + \Sigma_3 O$, respectively (Klein et al., 1992). Both of them involve phonon branches with very flat joint dispersion curves and possessing singularities in spectra. Different mechanisms may influence the density of states, but a plausible explanation of the observed spectral distortion is based on presence of hexagonal diamond polytypes. Comparison of densities of states for lonsdaleite (2H) and cubic (3C) diamond suggest that in the former phase intensities of discussed two-phonon bands should indeed be smaller than in 3C diamond (Salehpour and Satpathy, 1990).

One of the most interesting features of the yakutite IR spectra is the presence of strong $CO_2$ absorption. Persistence of this feature even in spectra acquired with purging of the spectrometer and sample compartment by dry nitrogen and, most importantly, shift of the band position and absence of rotational splitting clearly indicates that carbon dioxide is present as dry ice, $CO_2$-I. Similar bands were also observed in impact diamonds from Popigai (Koeberl et al. 1997) and in relatively rare terrestrial diamond single crystals (Schrauder and Navon, 1993). The $CO_2$ is obviously confined to nanoscale inclusions and the position of the symmetric stretch mode $\nu_3$ at 2360 cm$^{-1}$ indicates residual pressure of ~2 GPa following calibration proposed in (Hanson and Jones, 1981). In the case of Popigai impact diamonds Koeberl et al. (1997) suggested residual pressures of at least 5 GPa. Interestingly, examination of piezoeffects in Popigai nanodiamonds also suggest residual pressures in the range 2-4 GPa (Zilbershtein 1990), although this estimate is based on assumption of the absence of a contribution of lonsdaleite into diamond birefringence, which is probably incorrect (see Valter et al. 1992). Note, however, that in the spectra of our samples the bending mode $\nu_2$ at 658 cm$^{-1}$ shows only marginal Davydov splitting, thus indicating that the $CO_2$ ice in the inclusions is disordered and/or contains impurities (e.g. $H_2O$, $N_2$, etc). Subsequently, estimation of the residual pressure of the $CO_2$ inclusions from the spectroscopic data is ambiguous (Barannik et al. 2021). In addition, the quoted position of the bending $\nu_2$ mode almost coincide with the position at atmospheric pressure, thus being in contrast with the $\nu_3$ band. We note also that compressed carbon dioxide and water-containing inclusions may rupture poly- and nanocrystalline diamond during decompression as was demonstrated in Shiryaev et al. (2006).

*Photoluminescence spectroscopy*

Photoluminescence (PL) spectra shown in Fig. 5 are generally similar to those described in other studies of yakutites and impact diamonds in general (Kaminskii et al. 1985; Galimov et al. 1980; Yelisseev et al., 2016, 2018). In our work the PL spectra were recorded at room temperature, since for nanoparticulate or strongly strained systems cooling does not notably influence the width of spectral bands because of inhomogeneous, i.e. strain-related (e.g., Davies, 1970) line broadening. We do not observe common nitrogen- and/or vacancy defects typical to diamond single crystals, synthetic bulk nanodiamonds and impact diamonds from ureilites. The absence of such defects distinguishes yakutite from other impact diamonds. Yakutite PL spectra can be understood as a superposition of several bands; the relative intensity of these bands varies considerably both between the samples and within individual specimen without clear correlation with any other property or position in the sample, even on a surface exposed by polishing. The first broad and structureless band peaks at ~620 nm. A narrower band, which is likely a zero-phonon line (ZPL) with vibronic maxima, is located at 700 nm. Combined examination of spectra recorded both at 405 and 488 nm suggests that the former band actually consists of two components with maxima at ~590 nm and at 620 nm, the 620 nm maximum may be another ZPL. A weak band at 778 nm is also present.

At present models of the observed centers remain unknown, but for most of the observed features assignment to dislocation-related defects was proposed, namely, for 590-593 nm (Zaitsev 2001 and refs. therein) and for 620 nm (see Bokii et al. 1986 and refs. therein). The band at 700 nm is very likely identical to a band which is observed in the range 700-730 nm in many deformed and impact diamonds (Plotnikova et al. 1980; Galimov et al. 1980) as well as in ultra-pure synthetic nanocrystalline diamonds (Ikeda and Sumiya 2016). Initial assignment of this band to dislocation-related defects (Plotnikova et al., 1980) found some support in Mora et al. (2002, 2003). In a recent detailed investigation of PL spectra of deformed N-containing terrestrial diamonds (Hainschwang et al., 2020) it was suggested that the "700" nm band is a vibronic mode of a ZPL located at 620 nm, which, in turn, is clearly pronounced only at low temperatures. Whether the "700 nm" band from the later paper corresponds to features observed in yakutite or not is yet unknown, but in any case, its dislocation-related origin seems to be plausible. Taken together we tentatively assign most PL features seen in our set of yakutites to dislocation-related defects, possibly decorated by impurities.

### *X-ray tomography and radiography*

Absorption and phase-contrast modes of X-ray tomography were employed for investigation of the internal structure of yakutite samples (Fig. 6). In our previous study of a single yakutite specimen using these methods (Shiryaev et al., 2019) it was shown that in contrast to other cryptocrystalline natural diamonds such as carbonado containing numerous pores, cracks are the only noticeable defects in yakutite. We note that phase-contrast tomography is sensitive to cracks, grain boundaries and other similar imperfections even in cases of absence of the filling material. The present study confirms that although generally yakutites are dense materials with some cracks, pores may also exist. In some samples the pores are encountered in near-surface regions only, in the sample No. 406 they are present throughout the specimen (Fig. 6b).

However, for the purpose of the current study the most important result of the X-ray tomography is the observation of dense inclusions not connected to the surface by cracks (see Supplementary movie 1). In two of our samples several inclusions of roughly ellipsoidal shape are observed using tomography and radiography (Fig. 7). After their detection, X-ray radiography mapping was used to assess the composition of the inclusions *in situ*, i.e. without polishing or other ways to expose them. Examination of transmission images together with chemical maps shows that the only heavy element emitting X-rays with energies sufficient to penetrate through overlapping diamond are iron and zirconium. Iron clearly concentrates in cracks and mostly forms continuous domains (Fig. 7 first row), which almost certainly indicates secondary contamination. On the other hand, zirconium is observed in the inclusions only. At this stage it was yet impossible to identify the mineral, since thickness of the encompassing diamond was sufficiently large to suppress detection of silicon or another light element. In order to obtain information about the inclusions they were exposed by sample cutting, see next section.

### *Inclusions*

Using tomography and radiography data the inclusions were exposed by laser cutting and very gentle polishing to reduce heating. µ-XRF measurements of the exposed inclusions showed that they consist of Zr with noticeable (~2 mass% as oxide) of Hf (recall that only elements heavier than Na are detected). Note that hafnium is determined semi-quantitatively since the inclusions are too small to fulfil the condition of a semi-infinite sample required for correct evaluation of the concentration of elements with dramatically different energies of emission lines such as Zr and Hf. It is notable that no traces of silicon were detected. Raman spectroscopy was used to uniquely identify the inclusions as $ZrO_2$ (Fig. 8). Remarkably, besides expected baddeleyite bands, the spectrum contains several other broad peaks indicating an intimate mixture of monoclinic and tetragonal zirconia (m- and t-$ZrO_2$, respectively). Tetragonal zirconia is formed during high-temperature and/or high pressure treatment of the monoclinic polymorph and usually transforms back upon cooling. In the case of our samples the presence of t-$ZrO_2$ is a solid proof of high-temperature – high-pressure conditions, thus ruling out the possibility of secondary contamination.

Preservation of the tetragonal phase in quenched zirconia samples from shock experiments is possible at pressures exceeding 12 GPa (Morosin et al., 1984). It is well-known that t-$ZrO_2$ can be stabilized by the admixture of different elements, but no impurities were observed on exposed inclusions using µ-XRF except Hf, present in relatively low concentrations. The most plausible explanation of the persistence of t-$ZrO_2$ in the absence of impurities stabilizing high-temperature phase is a very small grain size of zirconia: it is known that nanometer-sized grains of this polymorph may survive in the stability field of a monoclinic phase, possibly, due to the highly strained structure of such grains (Garvie, 1965, 1978). Reducing conditions may also favor stabilization of t-$ZrO_2$ (see Shukla and Seal 2005 for a review). Tetragonal zirconia quenched from shock experiments may even survive heating by an electron beam if the grain size is sufficiently small (Hellmann et al., 1984).

## DISCUSSION

*Origin of single crystals in impact diamond*

Depending on the carbon precursor the mechanism of shock transformation of sp$^2$-C to diamond may be either martensitic (diffusion-less) for graphitic crystallites or diffusion–driven in case of amorphous carbons (Pyaternev et al., 1986, Kurdyumov et al. 2005). Orientation of the shock wave relative to the graphite basal planes strongly influences the minimal pressure required for the graphite-diamond transformation. In the case of Popigai, quartz and feldspar granoblasts may contain accumulations of differently oriented graphite scales (ch. 4.1.1. in Popigai impact structure…, 2019); consequently, the efficiency of the graphite-to-diamond transformation will vary even for flakes residing just few millimeters apart.

As suggested already in one of the first studies (Sokhor and Futergendler, 1975) and confirmed many times since (Valter et al., 1992, Popigai impact structure..., 2019 and refs. therein), impact diamonds mostly consist of nanocrystalline grains. Spectroscopic results of the current work, in particular, infra-red spectroscopy (see Figs. 4, 5 and relevant text), confirms the nanocrystalline character of the studied samples. At the same time, both optical microscopy and X-ray diffraction investigation of our samples indicates that impact-related yakutite consists of a nanocrystalline diamond matrix with embedded single crystal diamond grains reaching few tens of microns in size. To the best of our knowledge, the origin of micron-sized diamond grains in yakutites was not discussed in any detail and was only briefly mentioned by Kharlashina and Naletov (1991). It is thus important to discuss possible mechanisms of their formation and analyze data on other types of impact diamonds.

It is usually believed that the presumed short duration of the formation of impact diamonds precludes the formation of micron-sized single crystal grains. However, numerous reports of distinct spots due to diamond single crystals on X-ray patterns of impact diamonds of different provenance are known (e.g., Kharlashina and Naletov, 1991; Valter et al., 1992; Nestola et al., 2020). Single crystal diamond grains with sizes exceeding 1 μm were found in yakutite by Titkov et al. (2004) using transmission electron microscopy. Moreover, even laboratory experiments on shock compression of graphite may produce micron-sized grains giving single crystal electron diffraction patterns (Trueb, 1968); note that in these experiments the duration of the shock-induced transformation was very short in comparison with large scale meteoritic impacts. Amazingly, large facetted diamonds (tens of microns) were observed in shocked cast iron, for review see Sobolev et al. (1993). The authors of the later paper advance an idea of diffusion-driven post-shock growth of the diamonds from a highly supersaturated carbon solid solution. Although this mechanism seems doubtful to us, their observation deserves attention until a contamination or another similar fault in the experimental data of these authors is proven. Additional confirmation of the presence of micron-sized octahedral grains on the surfaces of natural impact diamonds was reported in a book by Valter et al. (1992) and set of papers by Kvasnitsa et al. (2015, 2016, 2018). Based on EPR and PL measurements these authors claim that the crystallites contain point defects, presumably nitrogen-related. However, from detailed examination of the papers and personal contacts with the authors we conclude that mentioned defects more likely reside not in the surface grains themselves, but rather they are present in diamond grains decorated by "single crystals".

Several processes may lead to the formation of relatively large crystalline diamond grains under impact. While in laboratory experiments the shock wave very rapidly leaves the sample volume, in the case of large scale events such as the Popigai impact, the compression stage was in the order of seconds (Popigai impact structure…, 2019). It is obvious that a seconds-long compression phase does not apply to the whole affected volume, but depending on position of the carbonaceous mineral in the target rock and thermomechanical and acoustic properties of surrounding phases, the pressure rise, its duration and magnitude vary in a very broad range. It is known that discharge of aqueous fluids from the overheated target rocks significantly slows down the rate of the pressure release (Kieffer and Simonds, 1980). According to Vishnevsky et al., (2006) in the Popigai event this mechanism is responsible for persistence of residual pressures of up to 3.3 GPa in tagamites for 10-12 seconds. One may suppose that even higher pressures could have been "preserved" in some locations for sufficient time required for formation of single diamond crystallites in yakutite samples.

Another possibility of formation of "large" diamond crystals in impact events was proposed by Kvasnitsa and co-authors (2015, 2016, 2018). According to these works, growth of morphologically perfect diamond crystallites on the surfaces of impact diamonds occurs by a chemical-vapour deposition (CVD) mechanism from the gas phase surrounding just-formed diamonds. The discussed crystals were observed on diamonds extracted by drilling of the impact structure and if the CVD mechanism is indeed responsible for their formation, the formation of a transient cavity rich in vaporized C-containing species must be inferred. Interestingly, the formation of a single crystal shell on a highly porous core is often observed in reactions of metasomatic mineral replacement with negative volume effect (Glikin 2009), which is similar to the graphite—diamond transition. Whereas yakutite formation cannot be directly compared with rather slow metasomatic processes, certain similarities may exist and the gas-phase precipitation in the transient void may fit this scenario. From the presence of tiny carbonaceous, presumably, diamond globules resembling ballas, Kvasnitsa (2018) suggests that micrometer-sized octahedral and other morphologically well-formed diamond crystals possibly grew by an oriented crystallite attachment mechanism. Whereas this mechanism may indeed be applicable, transformation of such an aggregate into a single crystal requires prolonged annealing.

Interestingly, detailed investigation of N and $^{36}$Ar content in yakutites (Shelkov et al. 1998) showed the presence of two gas carriers with thermochemical properties similar to diamond, with proportions varying both between individual samples and even between chips from the same yakutite specimens. In other words, the presence of two populations of diamond (or a diamond-like phase) is suggested. Based on the similarity of the Ar isotope ratio in one of the components with the atmospheric ones, Shelkov et al. (1998) also suggested that one of the populations may be related to CVD-type diamond growth from material vaporized by the impact.

### *ZrO$_2$ inclusions in yakutite*

Understanding of yakutite formation is complicated by a fact that these samples are observed in river placers only and their parent rocks are unknown. The majority of graphite in Popigai rocks is represented by flakes smaller than 2 mm, which are mostly found in various gneisses. Well-known small platy impact diamonds from Popigai were produced from such lamellae. Graphite segregations up to several centimeters in size occasionally observed as intergrowths in quartz and feldspars (ch. 4.1.1. in Popigai impact structure…, 2019) may be responsible for yakutite formation. However, the exact provenance of yakutite remains elusive.

Investigation of syngenetic inclusions is one of the most reliable ways to reconstruct the formation environment of a mineral. Inclusions of intimately mixed $ZrO_2$ polymorphs observed in our work is, to the best of our knowledge, the first observation of macroscopic inclusions in yakutite. The textural relationships between the trapped phase and diamond matrix indicates that $ZrO_2$ is inherited from a precursor carbonaceous phase. Therefore, understanding its origin may shed light on the local source of yakutite.

Information about inclusions in impact diamonds in general and in yakutites in particular is very limited and only few studies were published (Rumyantsev et al., 1981; Titkov et al., 2004; Ugap'eva et al., 2010, Yelisseyev et al., 2013). These inclusions are represented by graphite, various silicates, intermetallics, ilmenite and some other oxides. Whereas graphite may well be syngenetic and represent either incomplete transformation of the precursor or, in contrast, reverse transition of diamond on cooling, all other minerals are undoubtedly of secondary origin since they are present on grain surfaces, fill cracks and/or form ingrowths. The only really syngenetic inclusions known in impact diamonds are represented by $CO_2$ ice and, possibly, water, which were observed in IR spectra and, possibly, in TEM images (Koeberl et al. 1997). Analysis of trace elements in bulk impact diamonds by neutron activation (Shibata et al. 1993; Koeberl et al. 1997) and laser-ablation ICP (Yelisseyev et al., 2016b) revealed the presence of a noticeable amount of various impurities; REE patterns indicate high a similarity with crustal rocks (Shibata et al. 1993). However, in these studies it is very difficult to discriminate between secondary contamination and primary inclusions.

It is tempting to suggest zircon grains as a source of the $ZrO_2$, especially in view of reported finds of zircon grains in Popigai structure (e.g., Popigai impact structure…, 2019). It is well-known that zircon may decompose into $ZrO_2$ and $SiO_2$ during impact and products of zircon transformation are found in Popigai tagamites (Vishnevsky et al., 2006). However, despite a careful search, we have not detected any traces of Si-containing compounds in the inclusions neither by XRF nor by Raman, thus implying that baddeleyite and not zircon was present in the yakutite precursor. However, there are no reports of baddeleyite occurrence in Popigai thus complicating the explanation of our finds. As a tentative hypothesis we propose that the formation of at least some yakutite diamonds may be connected to carbonatite intrusions and/or pipes. In the south-east direction from the Popigai structure there are more than 300 carbonatite bodies belonging to the Chomurdakh, Orto-Yarginsky and Starorechensky fields of alkali-ultrabasic rocks (Marshintsev 1974, Entin et al. 1991). Accessory baddeleyite is present in all these bodies; its size varies from 0.3-0.5 to 1.5-2 mm. Both $ZrO_2$ and graphite are rather common in carbonatites (see, e.g., Kogarko and Ryabchikov, 2013) and baddeleyite inclusions in graphite were reported for carbonatite-containing rocks in India (Rajesh et al., 2006). Absence of similar inclusions in graphite from Yakutian carbonatites may well be explained by intense mechanical treatment employed for baddeleyite separation, which easily destroys graphitic carbon. Some of these carbonatite bodies are situated just ~20 km from the impact crater rim and the extent of the carbonatite region makes plausible hypothesis that some carbonatites were located within the crater and contributed to the formation of yakutite diamonds.

## CONCLUSIONS

Yakutite is a type of impact diamond, which is almost certainly related to the Popigai impact structure, or, less likely, to yet another unidentified crater. In this work results of the investigation of several well-preserved yakutite diamonds with contrasting morphology are presented. Peculiarities of their structure from the atomic to macroscopic scale are revealed and the most important findings are listed below.

1) Smearing of the two-phonon absorption bands in infra-red spectra of bulk nanocrystalline diamond is explained by the influence of small grain size and/or the presence of hexagonal diamond polytypes on phonon branches with flat joint dispersion curves. The one-phonon infra-red absorption is dominated by a band ascribed to stacking faults and other deformation-related defects. Similar defects are also responsible for photoluminescence spectra of yakutites.

2) The matrix of yakutite can be described as a bulk nanocrystalline diamond. However, in some yakutite samples diamond single crystals reaching several tens of microns are observed by optical microscopy; their crystallinity is confirmed by X-ray diffraction. These morphologically well-formed grains may reflect relatively long duration of a highly compressed state of rocks in impact events of the size of Popigai, or they might be a product of CVD-like growth from C-containing volatiles in the plume or in transient voids formed by shrinkage of the carbonaceous precursor during the graphite-to-diamond transition. Taken together with evidence from laboratory shock compression experiments (e.g., Trueb, 1968), the presence of micron-size single crystals cannot be used as an argument against the shock genesis of impact diamonds.

3) For the first time mineral protogenetic inclusions are observed in yakutite. These inclusions are represented by intimate mixture of monoclinic and tetragonal polymorphs of $ZrO_2$ with Hf admixture. Absence of $SiO_2$ traces excludes zircon, implying existence of baddeleyite in precursor graphite. Search for such target rocks may shed light on exact location of yakutite formation. Analysis of spectral features shows that syngenetic inclusions of $CO_2$ ice also contain water or another impurity (e.g., $N_2$), or represent highly disordered phase.

**Acknowledgements.**

Analytical measurements (XRF, Raman, PL) were performed using equipment of Center of Joint use of IPCE RAS. X-ray tomography studies were supported by RFBR grant 19-29-12035-mk. We highly appreciate comments and corrections made by Associate Editor, Dr. M. Poelchau, and by two reviewers.

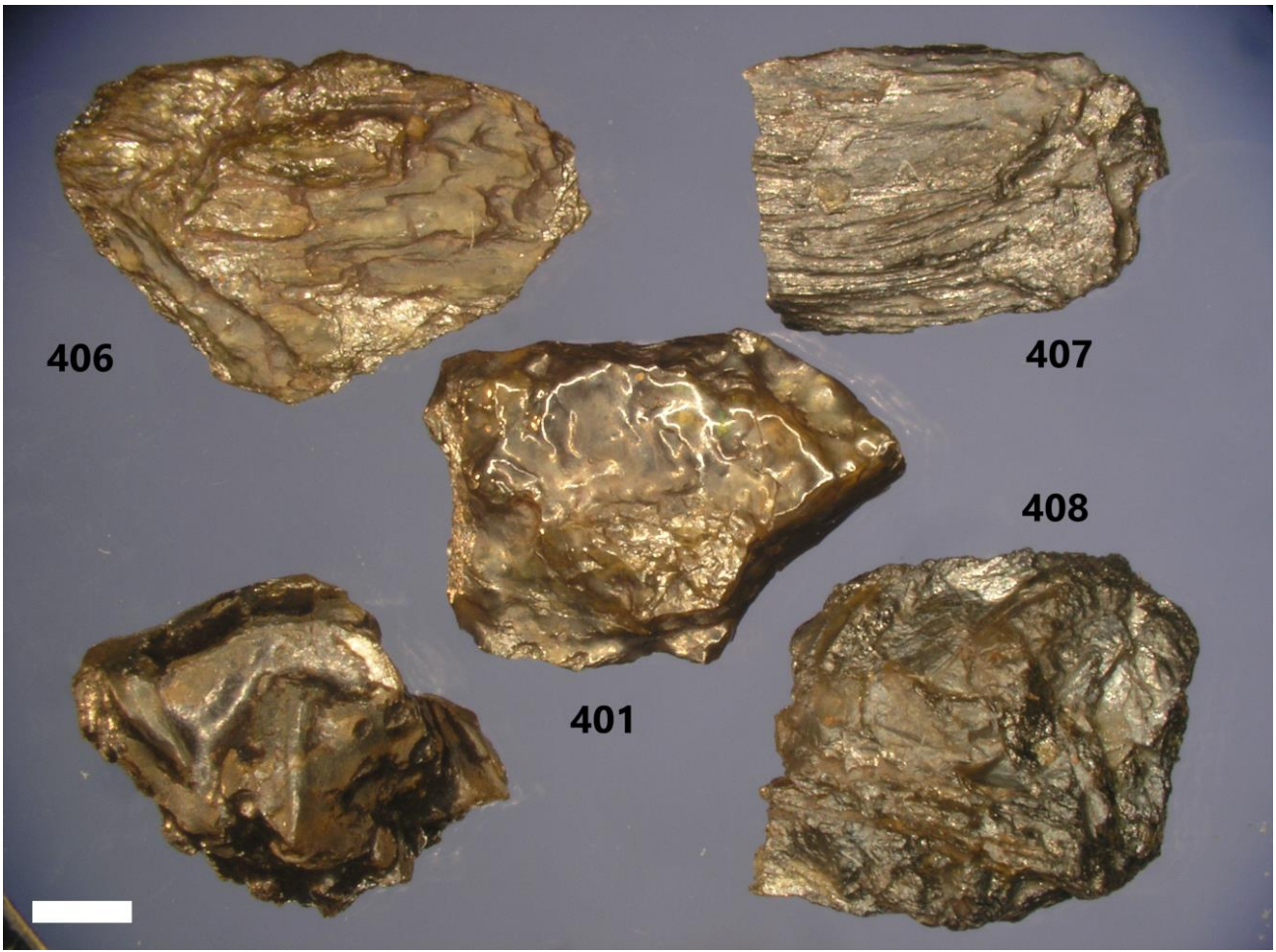

**Figure 1.** Optical photographs of yakutite samples. Scale bar – 1 mm.

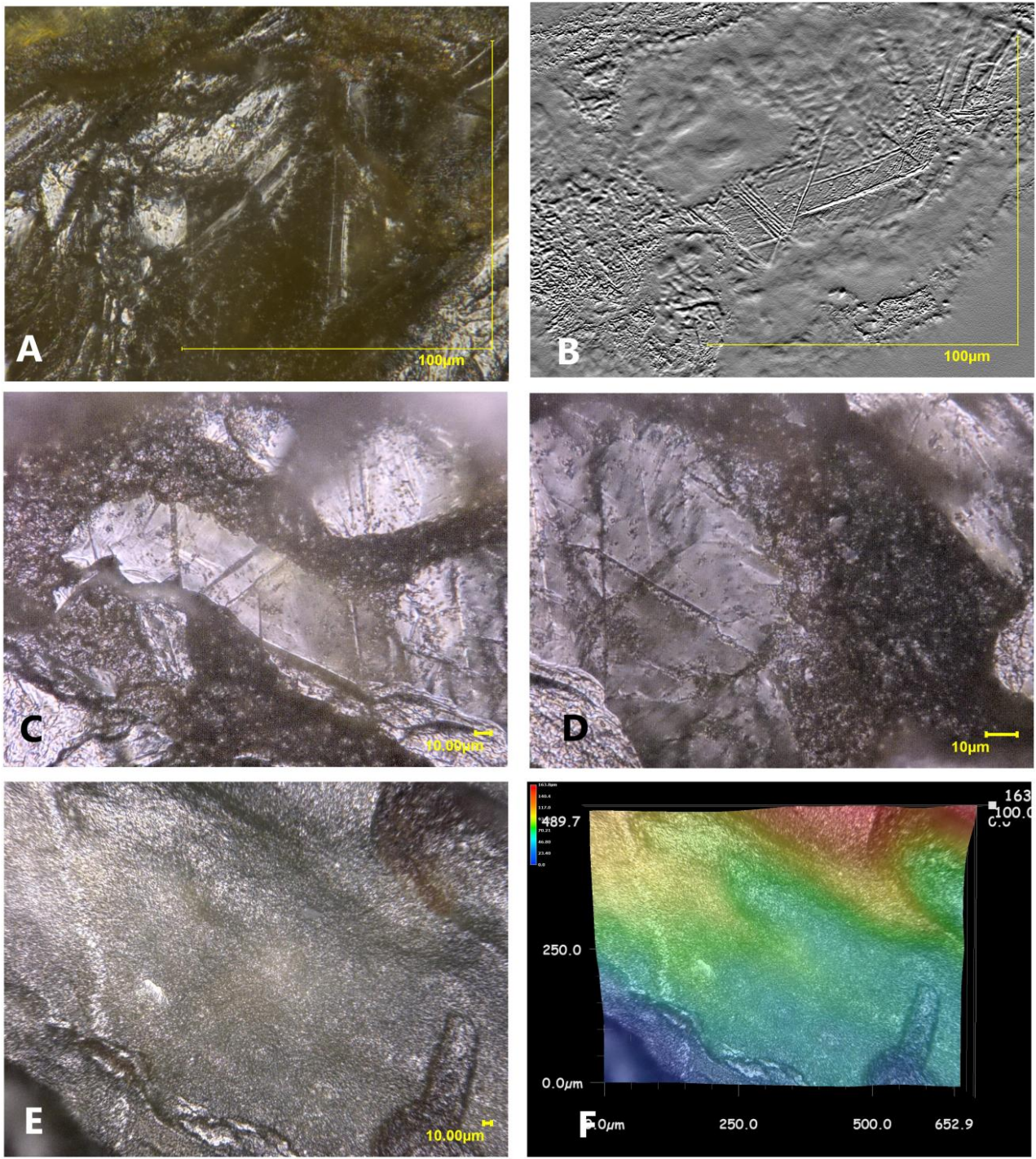

**Figure 2.** Optical micrographs of samples 407 and 408 obtained using 3D microscopy. Diamond single crystals embedded into nanostructures matrix are shown in A-D. B – image after filtering enhancing edges. E, F – face showing numerous submicron depressions due to oxidation, F – image E with 3D rendering.

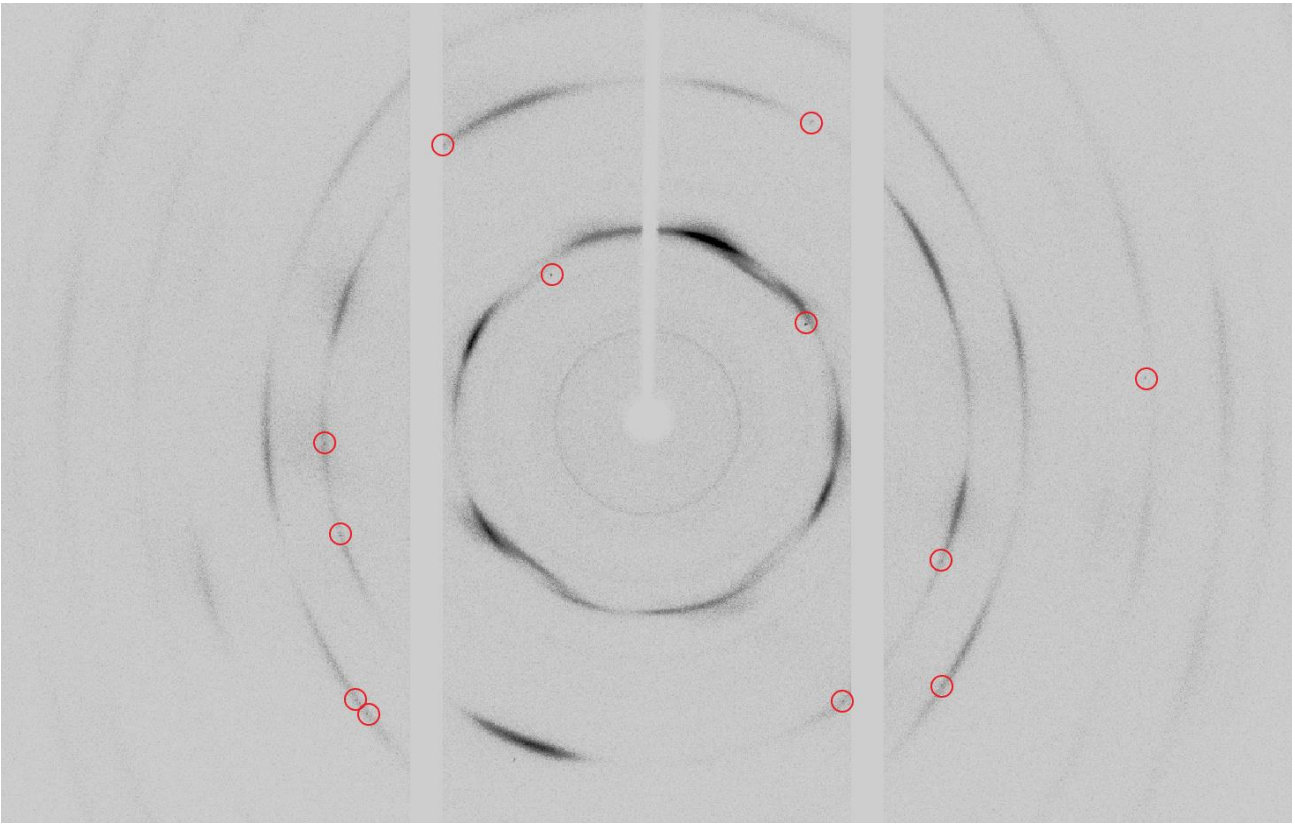

**Figure 3.** A typical frame of an X-ray 2D detector showing textured rings of diamond 111, 220 and 311 reflexes. Distinct single-crystal spots are marked with circles. Two vertical bands are due to contact between elements of the CCD detector.

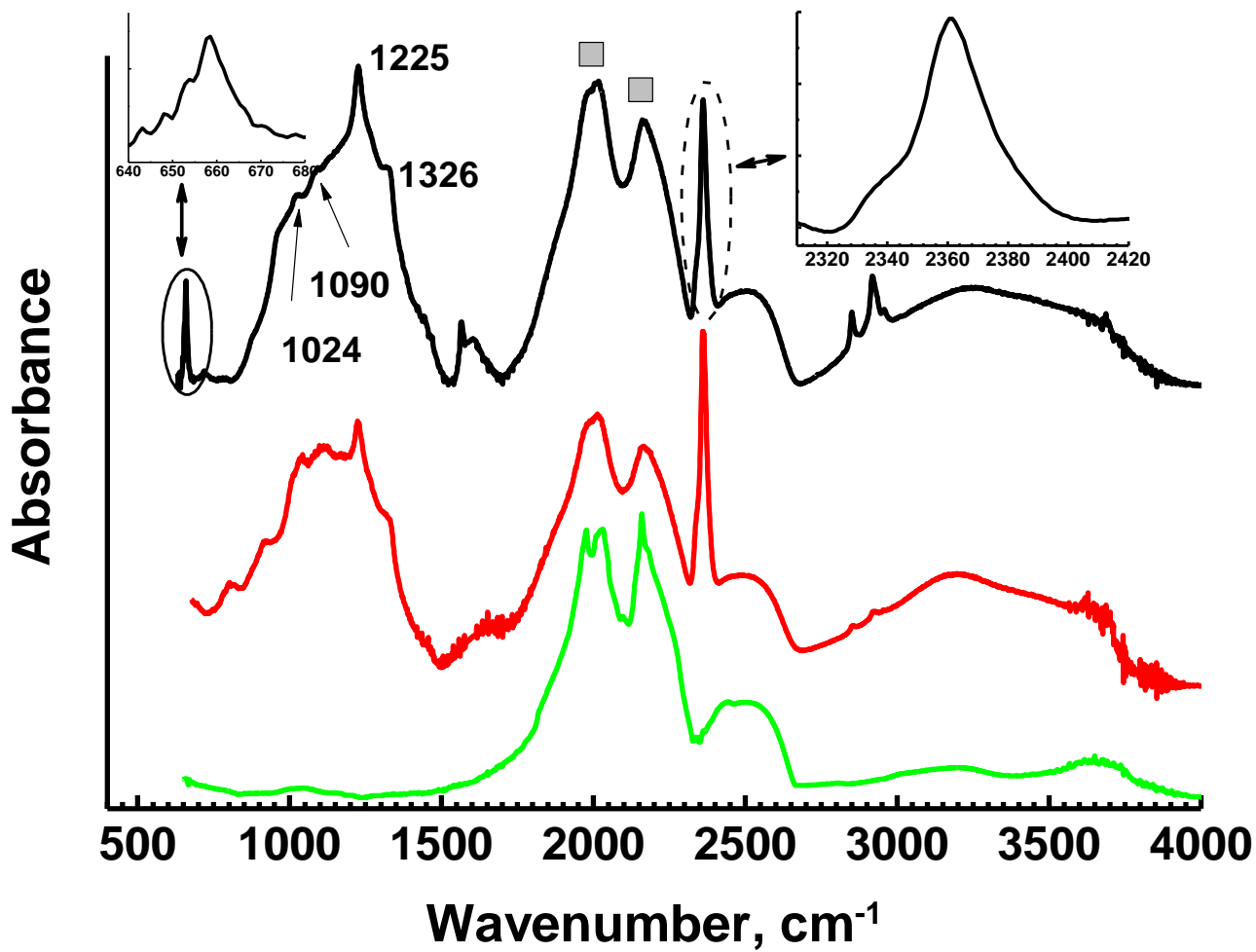

**Figure 4.** Representative Infra-red spectra of yakutite specimen (sample 407) in different zones. Spectrum of a synthetic type IIa diamond is shown for comparison (bottom curve). Filled squares show most prominent bands of diamond lattice absorption in two-phonon region. Insets show regions of $\nu_2$ and $\nu_3$ vibrations of $CO_2$. The curves are displaced vertically for clarity.

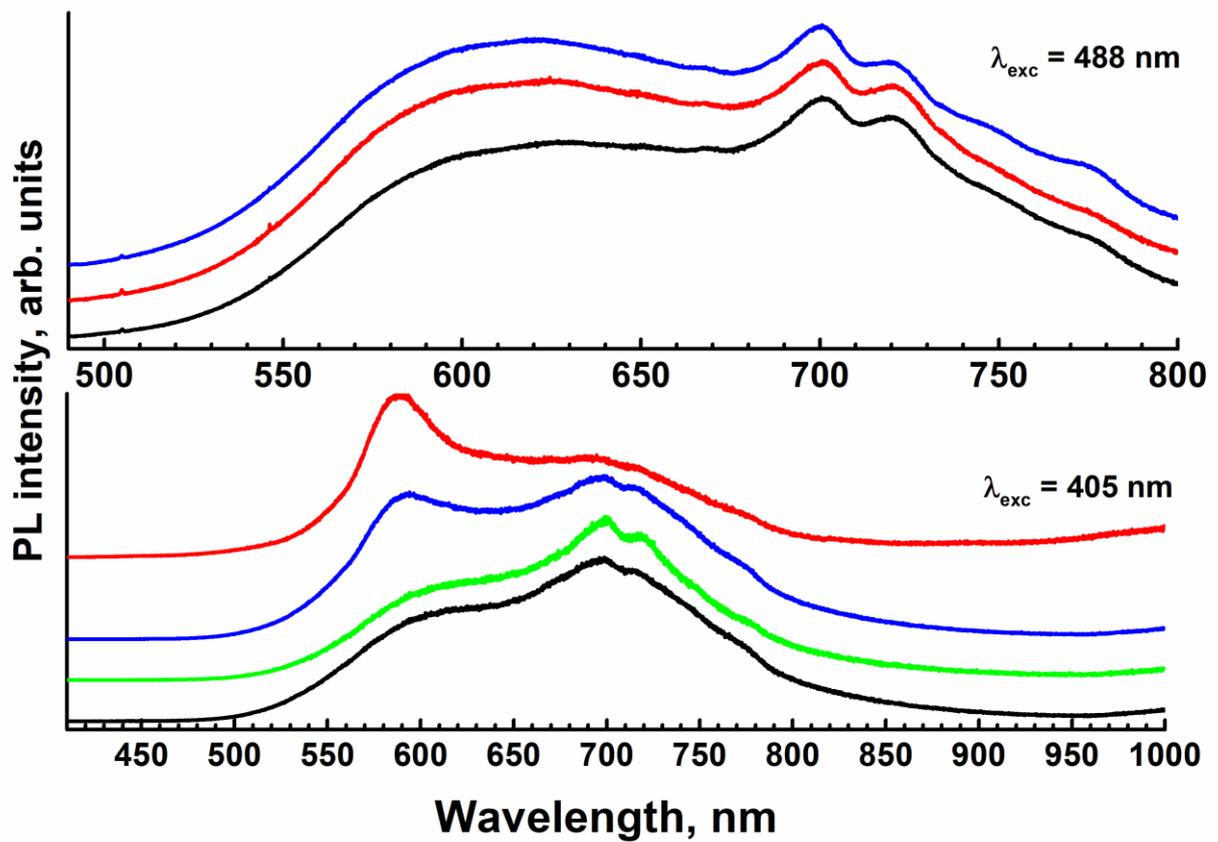

**Figure 5.** Photoluminescence spectra of yakutite samples recorded with excitation at 405 (bottom panel) and 488 nm (upper panel). The curves are displaced vertically for clarity; note differences in horizontal scale.

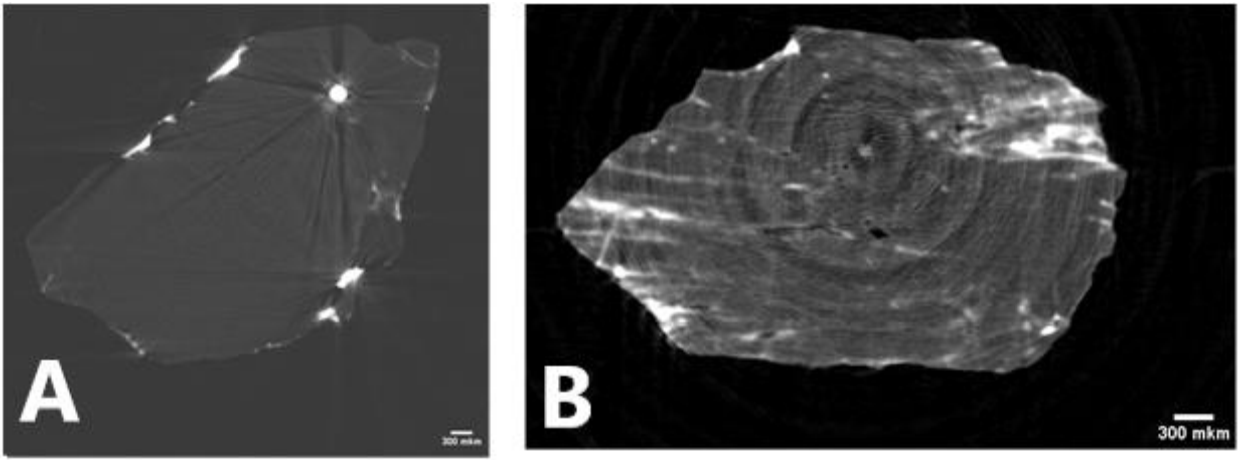

**Figure 6.** Selected projections of X-ray phase (A) and absorption (B) tomography. White spots correspond to inclusions containing highly absorbing phases.

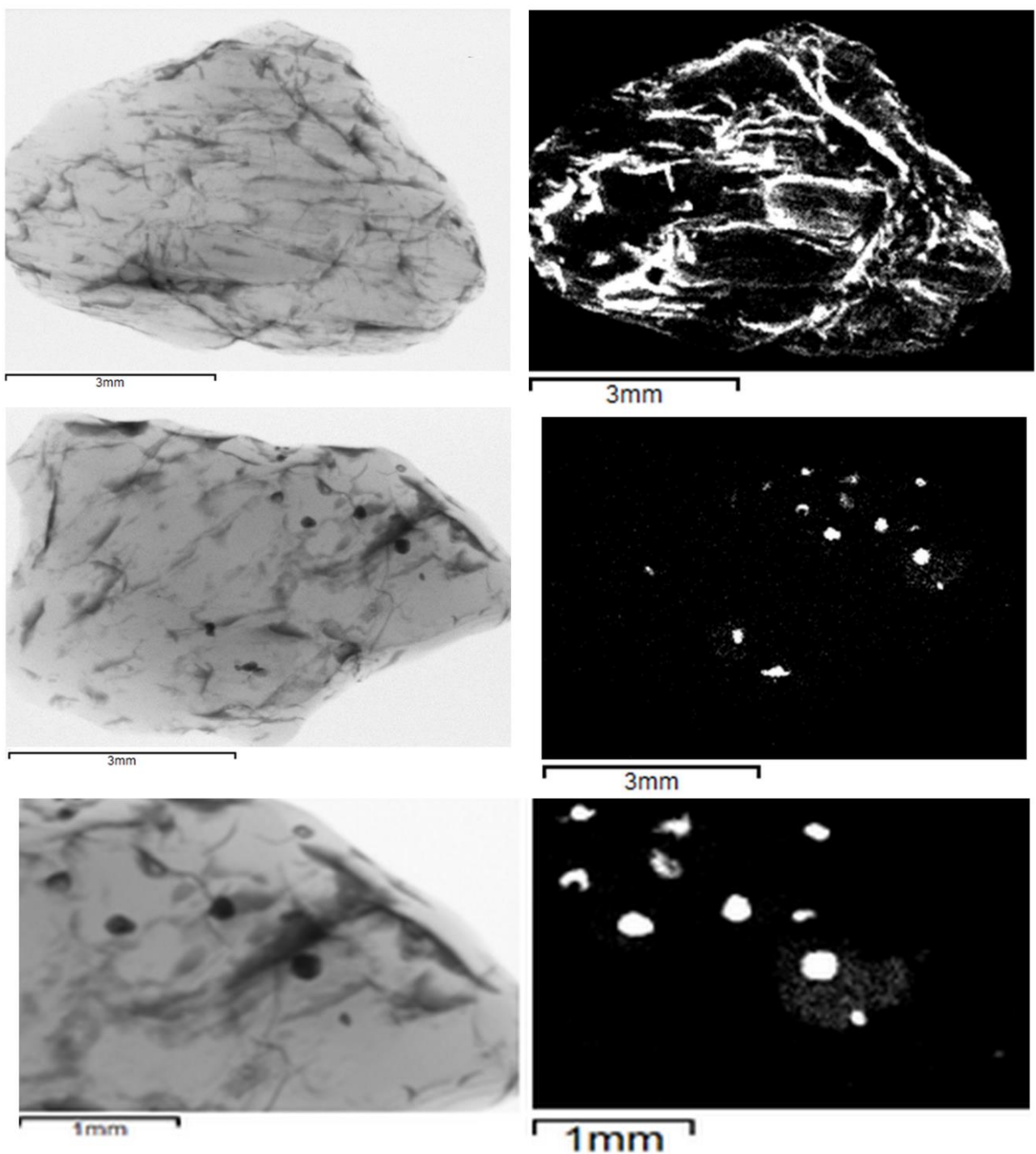

**Figure 7.** X-ray radiographic images (left column) and XRF maps (right column) of yakutite samples. Top row – sample 406, right – map of Fe Kα-line intensity. Second and third rows – sample 401. Right – map of Zr Kα-line intensity.

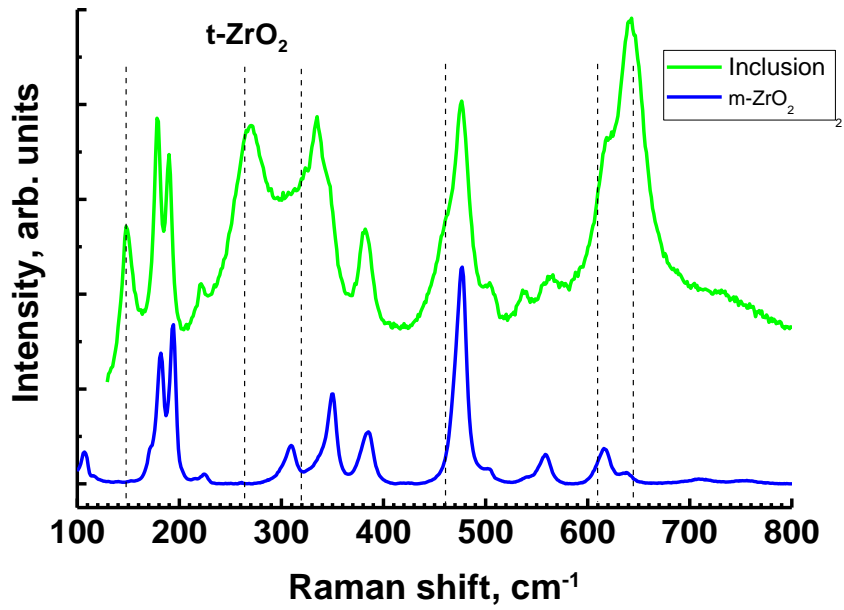

**Figure 8.** Raman spectrum of the ZrO$_2$ inclusion in sample 406 (green curve) and reference spectrum of baddeleyite (blue curve, Rruff ID R060078). Vertical lines show positions of raman peaks for tetragonal ZrO$_2$ modification (Kim et al., 1993). The curves are displaced vertically for clarity.